# GASTRIC SLOW WAVE MODELLING BASED ON STOMACH MORPHOLOGY AND NEURONAL FIRINGS

Tyas Pandu Fiantoro[1], Adhi Susanto[2], Bondhan Winduratna[3]

*Abstract*—Gastric content's mass and pH commonly assessed invasively using endoscopic biopsy, or semi-invasively using swallowable transducer. EGG (electrogastrography) is a technique for observing gastric myoelectrical activity non-invasively, that could be designed as mobile device.

In this research, 72 EGG recordings were obtained from 13 local white rabbit (*Oryctolagus cuniculus*). Recorded EGG processed using SCILAB 5.5.1 package. Signal processing consists of waveform identification altogether with recognition of resting, depolarization, ECA plateau, and repolarization segments of each EGG in the time domain based on amplitude and temporal filter. All rabbits were sacrificed after the recording in order to obtain its stomach content's mass and pH data. EGG waveform generator based on gastric morphological neuron assembly for local white rabbit (*Oryctolagus cuniculus*) modeled using those data.

If this model proved to be accurate, the mass and pH from rabbit (*Oryctolagus cuniculus*)'s stomach content could be assessed non-invasively, and could be a basis for human (*Homo sapiens*) trial.

[1] *Undergraduate, Electrical Engineering and Information Technology, Gadjah Mada University (eltroyaz@mail.ugm.ac.id)*
[2,3] *Departement of Electrical Engineering and Information Technology, Gadjah Mada University, Yogyakarta*

## I. INTRODUCTION

Electrogastrography (EGG) is a method for assessing gastric electrical activity that could be performed invasively (such as serosal EGG) or non-invasively (such as cutaneous EGG). More detailed description of EGG could be found in [1], [2], [3], [4], [5], [6], [7], [8], and [9]. This technique is unpopular compared to the ECG and EEG, because according to [10], physicians usually treat EGG as a research subject only. For the details of EGG assembly, measurement methods, and segmentation please refer to [11], and for the details of EGG parameterization please refer to [12]. This article only focus on the modelling of the obtained EGG waveforms.

## II. TERMS DEFINITION

### A. EGG

The abbreviation EGG may refer to electrogastrography (the technique to obtain gastric slow wave as mentioned by [2], [4], and [13]), electrogastrograph (the device that used to obtain the recordings), or electrogastrogram (the recorded signal). Hence EGG abbreviation should be interpreted appropriately based on the context.

### B. Biopotential

Biopotential is a portmanteau of bio-, which means related to life, and –potential, which means electrical potential. Biopotential emerged because of the ionic concentration difference among living systems such as cells, tissues, and organs.

### C. ICC

ICC is an abbreviation of Interstitial Cajal Cells, a collection of cells that found only in few digestive organs, including the stomach [14]. ICC classified as excitable cells, as they are generating action potential when exposed to ionic gradient environment [14]. The ICC stimuli consisted of neurotansmitters and hormons [15], [16]. Superposition of the electrical potentials in the ICC vicinity responsible for generating the gastric slow wave.

### D. Gastric Slow Wave

Commonly called "EGG waveform", its waveform shown in Fig. 1. It is constructed from resting potential segment (marked as "0"), depolarisation segment (marked as "1"), plateau segment (marked as "3"), and repolarisation segment (marked as "4"). Occasionally, spikes were present, the example of the spike could be seen in Fig. 1 as the region marked with "2".

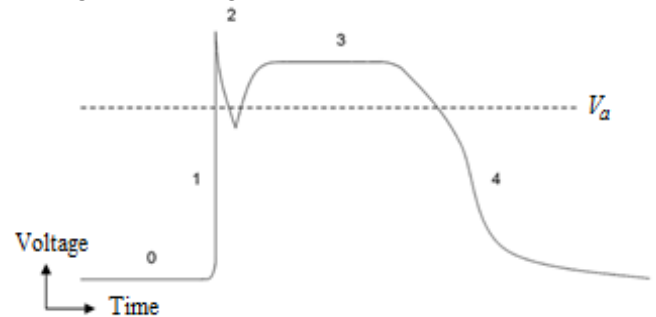

Fig. 1. One cycle of gastric slow wave according to [17] and [18]

## III. STOMACH MORPHOLOGICAL MODEL

The three dimensional shape of the stomach modeled using parametric equations. The idea begins from constructing a tube then distorting it according to its major and minor curvature. The tube starts with a circle with radius $r$ as its apex which centered at the transverse plane, $m\mathbf{i} + n\mathbf{j}$. Then, the same circle stacked cranially along $z$ to the $\mathbf{k}$ axis. To produce the major and minor curvature, the radius of each stacked circles is determined by a function of $z$, so that the "tube" has a largest radius in the *corpus* section, and going smaller to each *sphincter* (esophageal and pyloric). So far, the produced form still a straight tube with varied transverse radius, hence the centre of each transversal section must be arranged according to the $z$ value. Thus, to create the major and minor curvature, the centre of each transversal circle is determined by a function of $z$, namely a hypertrigonometrical function. The full parametric function to produce a three dimensional stomach morphology shown in Fig. 2 is described in (1).

$$GM(x, y, z, f, p, k_{cmaj}, k_{cmin}) = \left(\frac{k_{cmaj}e^{p-f}}{\cosh(z-p)\cosh(z-f)}\right.$$
$$\left. - k_{cmin}\frac{1 - \frac{1}{e^{z-f}+1}}{e^{z-p}+1}\right)\cos(\theta)\mathbf{i} + \left(\left(\frac{k_{cmaj}e^{p-f}}{\cosh(z-p)\cosh(z-f)}\right.\right.$$
$$\left.\left. - k_{cmin}\frac{1 - \frac{1}{e^{z-f}+1}}{e^{z-p}+1}\right)\sin(\theta) + k_{cmin}\frac{1 - \frac{1}{e^{z-f}+1}}{e^{z-p}+1}\right)\mathbf{j} + z\mathbf{k} \quad (1)$$

With *GM* is the generating function to produce the three dimensional stomach model, $k_{cmaj}$ is the eccentricity of the major curvature, $k_{cmin}$ is the eccentricity of the minor curvature, $f$ is the fundus coordinate, $p$ is the pylorus coordinate, $\theta$ is the angle domain for transverse plane (always from 0 to $2\pi$), $z$ is the axial domain for the length of longitudinal axis (always caudal, from $f$ to $p$), $\mathbf{i}$ is the transverse axis, $\mathbf{j}$ is the anteroposterior axis, and $\mathbf{k}$ is the longitudinal axis.

$$l_p = \int_f^p \sqrt{\left(\frac{\partial}{\partial z}k_{cmin}\frac{1 - \frac{1}{e^{z-f}+1}}{e^{z-p}+1}\right)^2 + 1} \quad (2)$$

This parametric function based morphological modelling proved useful not only for visualisation purpose, but also for calculating the length of the muscle fibre and its corresponding neuronal lengths. The length of a curve ($l_p$) that follows each cross sectional

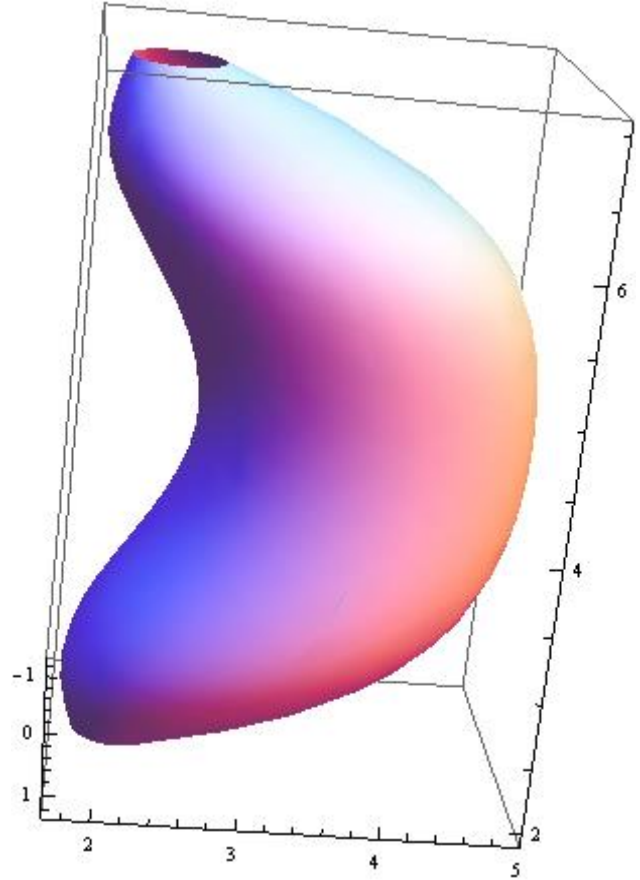

Fig. 2. Generated stomach with $f = 2$, $p = 7$, $k_{cmaj} = \frac{1}{4}$, $k_{cmin} = \frac{7}{8}$

(transverse) centre of the stomach caudally, expressed in curve equation (2).

## IV. STOMACH PHYSIOLOGICAL MODEL

The stomach assumed to be having three distinct phases, those are cephalic phase, gastric phase, and intestinal phase. This assumption is based on [14].

At cephalic phase, the stomach modeled to produce a short (5 to 10 seconds period) but constant gastric slow wave. Gastric mass assumed low (zero for initial condition, but nonzero for residual that appear in recurrent phase), and its pH also low (0 to 2). The stomach owner may feeling hungry, or being desired to eat at this phase.

At gastric phase, the owner of the stomach assumed had eaten certain mass of food. Hence, its gastric mass and pH is roughly equal to the initial eaten food. The stomach now actively digesting the foods, its pH is decreasing along with the fluctuation of the gastric mass.

At intestinal phase, the gastric mass modeled with the leftover mass called the residuals. Most of the food consumed had delivered to the jejunum or further. Its pH is increasing because the gastric chief cells stopped secreting gastric acid.

From the three phasic stomach, the average amplitude value of the plateau segments, $A(m,pH)$ modelled as (3).

$$A(m, pH) = 55\left(e^{(1.4-pH)^2 + \left(0.946 - \frac{m}{50}\right)^2} - 0.927\right) \quad (3)$$

The comparison of the plot of (3) with the observed data from [12], presented in Fig 3.

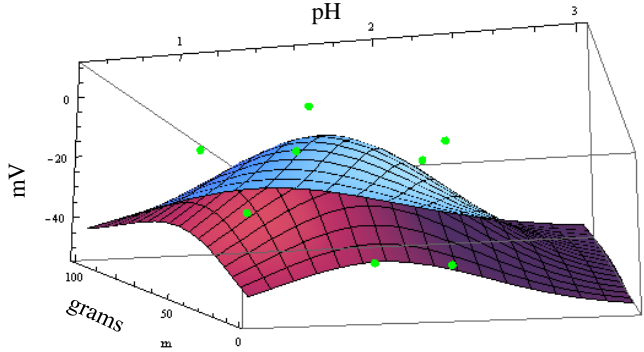

Fig. 3. 3D plot from (3) versus the dots of observed data from [12]

The duration of each plateau segment, $T_{1/2}(m,pH)$ given in (4), is also modeled with the three phasic concept.

$$T_{1/2}(m, pH) = \frac{\left(1 - \left(1 + e^{\frac{m-40}{6}}\right)^{-1}\right)\left(1 - \left(1 + e^{1.403pH - 1.5}\right)^{-1}\right)}{0.02325\left(1 + e^{0.181(m-74)}\right)\left(1 + e^{pH-9}\right)} \quad (4)$$

The comparison of the three dimensional plot of (4) with the observed data from [12], could be seen in Fig 4.

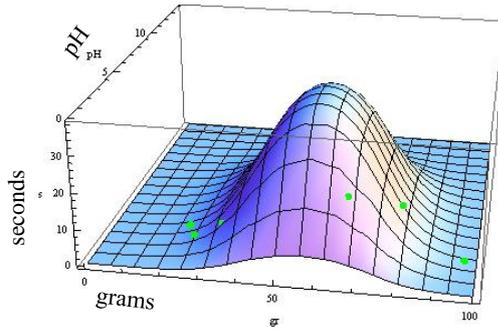

Fig. 4. 3D plot from (4) versus the dots of observed data from [12]

V. NEURONAL MODEL

Gastric ICC assembly modeled with the classical cable theory. Its model given in (5) with the initial function at $x = 0$ is $V_{ICC}$, which is solved numerically using SCILAB.

$$\lambda \frac{\partial^2 V}{\partial x^2} = \tau \frac{\partial V}{\partial t} + V \quad (5)$$

The $V_{ICC}$ function is a firing neuron model based on Nam and Miftahof observation. The mentioned function consisted of 18 stages algebraic non-linear differential equations, full details could be found in [19]. This $V_{ICC}$ function is coupled with the potassium, sodium, calcium, calcium-potassium, and chlorine ion channels, and it also depends on the $Ca^{2+}$ concentration dynamics. This $V_{ICC}$ function solved numerically with SCILAB 5.5.1 to produce a $V_{ICC}$ function as seen in Fig. 5. Each ion channel's current function presented in Fig. 6. Other channel dynamics, including the $[Ca^{2+}]$ dynamics could be viewed in Fig. 7 and Fig. 8.

Gastric slow wave generation modeled along with the migrating motor complex (MMC) pathway. The MMC trajectory modeled as the moving pulsating source, which follows the path $\zeta$ according to (6).

$$\zeta(t, m, pH) = \frac{(\tanh(t)+1)(\tanh(T_{1/2}(m, pH) - t) + 1)}{-13.0828 \log(0.5357 - 0.00397 A(m, pH))} \quad (6)$$

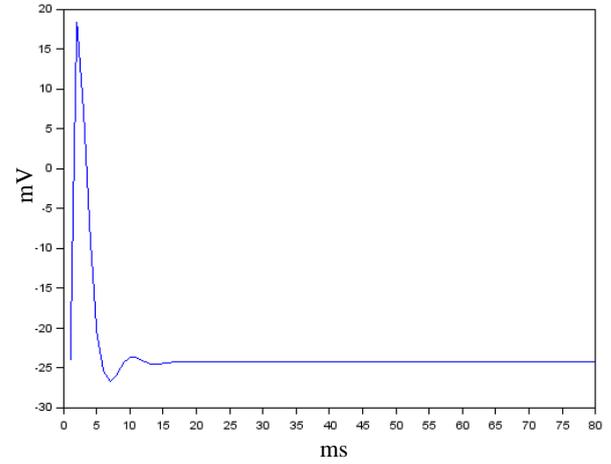

Fig. 5. The initial neuron firing function, $V_{ICC}$

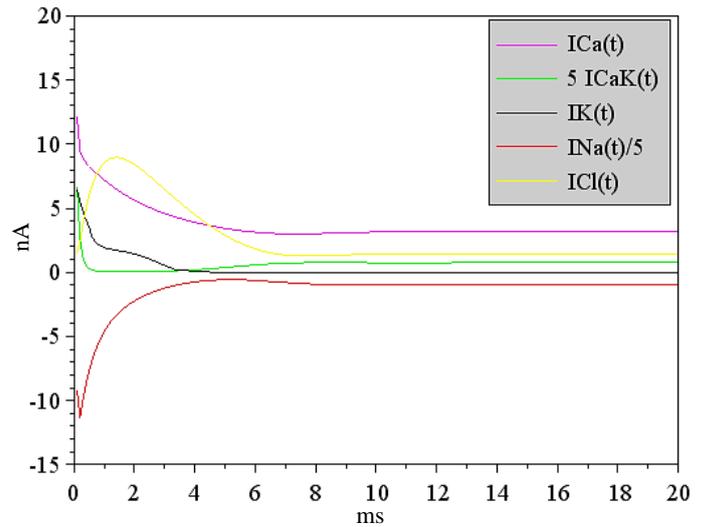

Fig. 6. Each modeled ion channel current dynamics

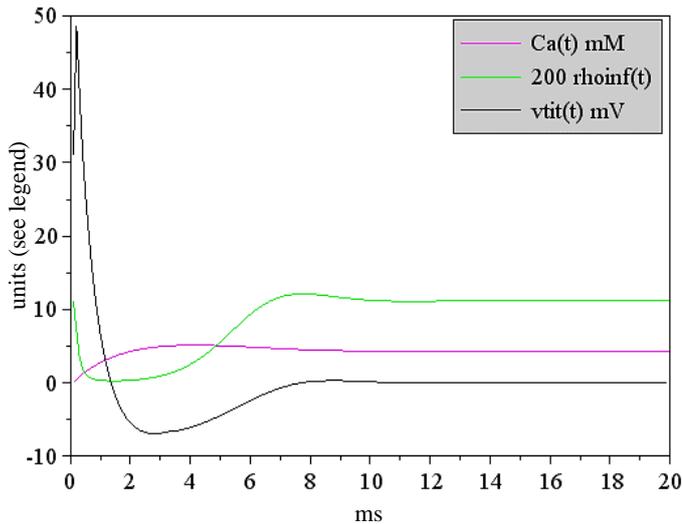

Fig. 7. Ca²⁺ concentration (magenta), its gate probability (green) and the first differential of the ICC pulse (black) dynamics

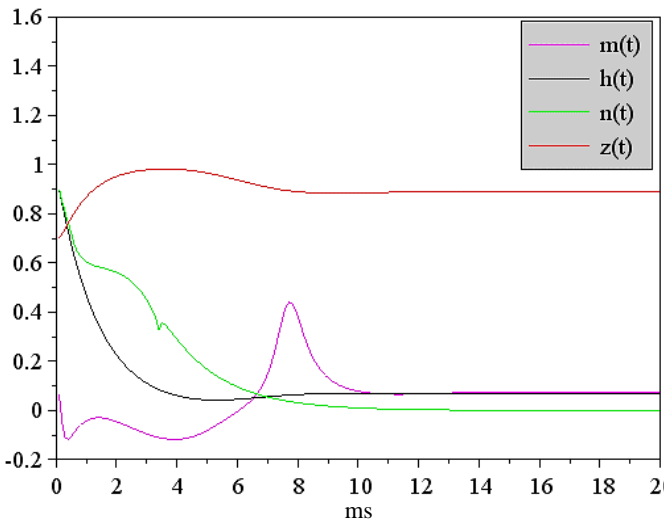

Fig. 8. Each modeled ion channel's gating coefficient

Whenever there are ICC activities, the potential at arbitrary point $q$ at the vicinity could be observed. This electric potential strongly affected by the pulsating source described at (6). Hence, the potential at the $q$ defined as the sum of each contributing pulsating source during its pulsating period. This could be expressed in time integral as in (7), where $V$ is the solution of (5).

$$V_q = \int V(|q - \zeta(\tau, m, pH)|, t - \tau)\, \partial\tau \tag{7}$$

The fundus is defined as the zero point of the gastric curve measurement ($q = 0$), hence (7) become (8) for the fundic potential value.

$$V_f = \int V(\zeta(\tau, m, pH), t - \tau)\, \partial\tau \tag{8}$$

It follows from (2) that the pylorus is located at the curved distance $l_p$ from the fundus. Substituting $q = l_p$, equation (9) was derived.

$$V_p = \int V(l_p - \zeta(\tau, m, pH), t - \tau)\, \partial\tau \tag{9}$$

Hence, the final difference between the fundic point and the pyloric point expressed as (10). This potential difference defined as the gastric slow wave potential.

$$V_{EGG} = V_f - V_p \tag{10}$$

## VI. DISCUSSION

The model for generating the slow gastric wave was able to mimic the actual observed data. An example of the comparison between modeled wave versus the actual observed wave is shown in Fig. 9.

The red curve in Fig. 9. was generated by using a mass value between 95 gr and 100 gr, and a pH value between 1 and 1.1 as the argument of the trajectory function $\zeta(t, m, pH)$. Modelling parameters of $R_a = 100$ Ω.cm, $r_m = 4900$ Ω/cm, and $c_m = 1$ μF/cm was used for the $V_{ICC}$ generator.

The green curve in Fig. 9 was an EGG recording obtained from a rabbit (*O. cuniculus*) that was found to be having a gastric content's mass and pH of 97.165 gr and 1.07 respectively. The root mean square error is 6.83 mV.

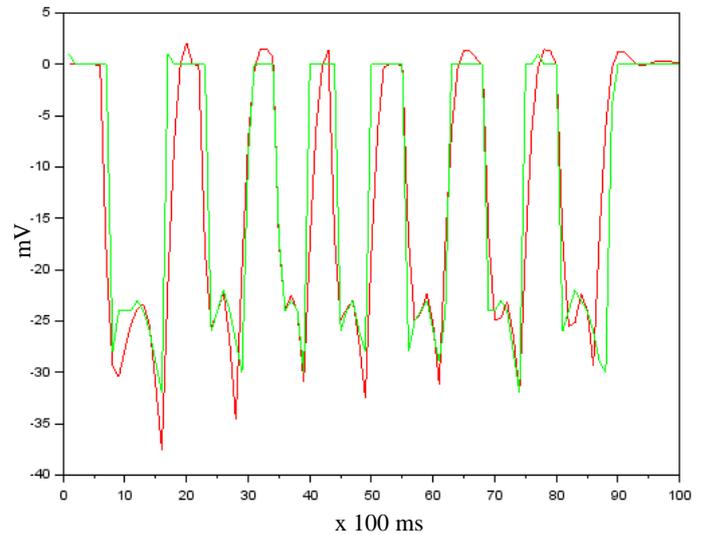

Fig. 9. The modeled EGG (red) versus the actual EGG (green)

The physical gastric parameters are not constants, hence their dynamics must be recorded in real time. Unfortunately, due to the technical limitations, the gastric physical values obtained in this research were based on "final" state, that is after the rabbits sacrificed. In short, sophisticated EGG modelling must be achieved in order to replace the common gastric invasive assessment.